\documentclass[a4paper,11pt]{article}
\pdfoutput=1 % if your are submitting a pdflatex (i.e. if you have
\usepackage{jcappub} % for details on the use of the package, please
\usepackage{multirow}
\usepackage{svg}
\usepackage[T1]{fontenc} % if needed
\usepackage[normalem]{ulem}
\usepackage{graphicx}
\usepackage{txfonts}
\usepackage{xcolor}
\usepackage{gensymb}
\usepackage{natbib}
\usepackage{lineno}
\usepackage{comment}
\usepackage{hyperref}
\PassOptionsToPackage{hyphens}{url}

\def\healpix{\texttt{HEALPix}}
\def\healpy{\texttt{healpy}}
\def\nside{$N_{\mathrm{side}}$}
\def\lb{\textit{LiteBIRD}}

 \title{First release of \lb\ simulations from an end-to-end pipeline.}

\author[1,2]{M.\,Bortolami,}
\author[1]{N.\,Raffuzzi,}
\author[1,2,3]{L.\,Pagano,}
\author[4,5,6]{G.\,Puglisi,}
\author[7]{A.\,Anand,}
\author[8]{A.\,J.\,Banday,}
\author[2,9,10]{P.\,Campeti,}
\author[1,7]{G.\,Galloni,}
\author[11]{A.\,I.\,Lonappan,}
\author[12]{M.\,Monelli,}
% ...
\affiliation[1]{Dipartimento di Fisica e Scienze della Terra, Università di Ferrara, Via Saragat 1, 44122 Ferrara, Italy}
\affiliation[2]{INFN Sezione di Ferrara, Via Saragat 1, 44122 Ferrara, Italy}
\affiliation[3]{Université Paris-Saclay, CNRS, Institut d’Astrophysique Spatiale, 91405, Orsay, France}
\affiliation[4]{Dipartimento di Fisica e Astronomia, Universitá degli Studi di Catania, Via S. Sofia,64, 95123, Catania, Italy}
\affiliation[5]{INAF, Osservatorio Astrofisico di Catania, via S.Sofia 78, I-95123 Catania, Italy}
\affiliation[6]{INFN, Sezione di Catania, via S.Sofia 64, I-95123, Catania, Italy}
\affiliation[7]{Dipartimento di Fisica, Università di Roma Tor Vergata, Via della Ricerca Scientifica, 1, 00133, Roma, Italy}
\affiliation[8]{IRAP, Université de Toulouse, CNRS, CNES, UPS, Toulouse, France}
\affiliation[9]{Max Planck Institute for Astrophysics, Karl-Schwarzschild-Str. 1, D-85748 Garching, Germany}
\affiliation[10]{Excellence Cluster ORIGINS, Boltzmannstr. 2, 85748 Garching, Germany}
\affiliation[11]{University of California, San Diego, Department of Physics, San Diego, CA 92093-0424, USA}
\affiliation[12]{Kavli Institute for the Physics and Mathematics of the Universe (Kavli IPMU, WPI), UTIAS, The University of Tokyo, Kashiwa, Chiba 277-8583, Japan}
% ...

   \date{Received \today }

   \abstract{}{}{}{}{} 
 

\affiliation[a]{One University,\\some-street, Country}
\affiliation[b]{Another University,\\different-address, Country}
\affiliation[c]{A School for Advanced Studies,\\some-location, Country}

\emailAdd{marco.bortolami@unife.it}
\emailAdd{nicolelia.raffuzzi@unife.it}
\emailAdd{luca.pagano@unife.it}
\emailAdd{giuseppe.puglisi2@unict.it}

\abstract{The \lb\ satellite mission aims at detecting  Cosmic Microwave Background $B$~modes with unprecedented precision, targeting a total error on the tensor-to-scalar ratio $r$ of $\delta r \sim 0.001$. Operating from the L2 Lagrangian point of the Sun--Earth system, \lb\ will survey the full sky across 15 frequency bands (34 to 448 GHz) for 3 years.The current \lb\ baseline configuration employs 4508 detectors
sampling at 19.1 Hz to achieve an effective polarization sensitivity
of $2\ \mu\mathrm{K\,arcmin}$ and an angular resolution of 31 arcmin (at 140 GHz).
 
We describe the first release of the official \lb\ simulations, realized with a new simulation pipeline developed using the \lb\ Simulation Framework. This pipeline generates 500 full-sky simulated maps at a \healpix\ resolution of \nside=512.   
The simulations include also one year of Time Ordered Data (TOD) for approximately one-third of \lb’s total detectors.}

\def\setsymbol#1#2{\expandafter\def\csname #1\endcsname{#2}}
\def\getsymbol#1{\csname #1\endcsname}

\def\Planck{\textit{Planck}}

\newbox\tablebox    \newdimen\tablewidth
\def\leaderfil{\leaders\hbox to 5pt{\hss.\hss}\hfil}
\def\tablenote#1 #2\par{\begingroup \parindent=0.8em
    \abovedisplayshortskip=0pt\belowdisplayshortskip=0pt
    \noindent
    $$\hss\vbox{\hsize\tablewidth \hangindent=\parindent \hangafter=1 \noindent
    \hbox to \parindent{$^#1$\hss}\strut#2\strut\par}\hss$$
    \endgroup}

\def\L2{\ifmmode L_2\else $L_2$\fi}

\def\DeltaT{\ifmmode \Delta T\else $\Delta T$\fi}
\def\deltat{\ifmmode \Delta t\else $\Delta t$\fi}
\def\fknee{\ifmmode f_{\rm knee}\else $f_{\rm knee}$\fi}
\def\Fmax{\ifmmode F_{\rm max}\else $F_{\rm max}$\fi}
\def\solar{\ifmmode{\rm M}_{\mathord\odot}\else${\rm M}_{\mathord\odot}$\fi}
\def\Msolar{\ifmmode{\rm M}_{\mathord\odot}\else${\rm M}_{\mathord\odot}$\fi}
\def\Lsolar{\ifmmode{\rm L}_{\mathord\odot}\else${\rm L}_{\mathord\odot}$\fi}
\def\inv{\ifmmode^{-1}\else$^{-1}$\fi}
\def\mo{\ifmmode^{-1}\else$^{-1}$\fi}
\def\sup#1{\ifmmode ^{\rm #1}\else $^{\rm #1}$\fi}
\def\expo#1{\ifmmode \times 10^{#1}\else $\times 10^{#1}$\fi}
\def\,{\thinspace}
\def\lsim{\mathrel{\raise .4ex\hbox{\rlap{$<$}\lower 1.2ex\hbox{$\sim$}}}}
\def\gsim{\mathrel{\raise .4ex\hbox{\rlap{$>$}\lower 1.2ex\hbox{$\sim$}}}}

\def\simprop{\mathrel{\raise .4ex\hbox{\rlap{$\propto$}\lower 1.2ex\hbox{$\sim$}}}}
\def\deg{\ifmmode^\circ\else$^\circ$\fi}
\def\pdeg{\ifmmode $\setbox0=\hbox{$^{\circ}$}\rlap{\hskip.11\wd0 .}$^{\circ}
          \else \setbox0=\hbox{$^{\circ}$}\rlap{\hskip.11\wd0 .}$^{\circ}$\fi}
\def\arcs{\ifmmode {^{\scriptstyle\prime\prime}}
          \else $^{\scriptstyle\prime\prime}$\fi}
\def\arcm{\ifmmode {^{\scriptstyle\prime}}
          \else $^{\scriptstyle\prime}$\fi}
\newdimen\sa  \newdimen\sb
\def\parcs{\sa=.07em \sb=.03em
     \ifmmode \hbox{\rlap{.}}^{\scriptstyle\prime\kern -\sb\prime}\hbox{\kern -\sa}
     \else \rlap{.}$^{\scriptstyle\prime\kern -\sb\prime}$\kern -\sa\fi}
\def\parcm{\sa=.08em \sb=.03em
     \ifmmode \hbox{\rlap{.}\kern\sa}^{\scriptstyle\prime}\hbox{\kern-\sb}
     \else \rlap{.}\kern\sa$^{\scriptstyle\prime}$\kern-\sb\fi}
\def\ra[#1 #2 #3.#4]{#1\sup{h}#2\sup{m}#3\sup{s}\llap.#4}
\def\dec[#1 #2 #3.#4]{#1\deg#2\arcm#3\arcs\llap.#4}
\def\deco[#1 #2 #3]{#1\deg#2\arcm#3\arcs}
\def\rra[#1 #2]{#1\sup{h}#2\sup{m}}

\def\dots{\relax\ifmmode \ldots\else $\ldots$\fi}
\def\WHzsr{\ifmmode $W\,Hz\mo\,sr\mo$\else W\,Hz\mo\,sr\mo\fi}
\def\mHz{\ifmmode $\,mHz$\else \,mHz\fi}
\def\GHz{\ifmmode $\,GHz$\else \,GHz\fi}
\def\mKs{\ifmmode $\,mK\,s$^{1/2}\else \,mK\,s$^{1/2}$\fi}
\def\muKs{\ifmmode \,\mu$K\,s$^{1/2}\else \,$\mu$K\,s$^{1/2}$\fi}
\def\muKRJs{\ifmmode \,\mu$K$_{\rm RJ}$\,s$^{1/2}\else \,$\mu$K$_{\rm RJ}$\,s$^{1/2}$\fi}
\def\muKHz{\ifmmode \,\mu$K\,Hz$^{-1/2}\else \,$\mu$K\,Hz$^{-1/2}$\fi}
\def\MJysr{\ifmmode \,$MJy\,sr\mo$\else \,MJy\,sr\mo\fi}
\def\MJysrmK{\ifmmode \,$MJy\,sr\mo$\,mK$_{\rm CMB}\mo\else \,MJy\,sr\mo\,mK$_{\rm CMB}\mo$\fi}
\def\microns{\ifmmode \,\mu$m$\else \,$\mu$m\fi}

\def\muK{\ifmmode \,\mu$K$\else \,$\mu$\hbox{K}\fi}
\def\microK{\ifmmode \,\mu$K$\else \,$\mu$\hbox{K}\fi}
\def\muW{\ifmmode \,\mu$W$\else \,$\mu$\hbox{W}\fi}
\def\kms{\ifmmode $\,km\,s$^{-1}\else \,km\,s$^{-1}$\fi}
\def\kmsMpc{\ifmmode $\,\kms\,Mpc\mo$\else \,\kms\,Mpc\mo\fi}
\providecommand{\sorthelp}[1]{}

\begin{document}
\maketitle
\flushbottom
\section{Introduction}\label{sec:introduction}
The standard cosmological model describing our Universe, referred to as the   $\Lambda$ Cold Dark Matter ($\Lambda$CDM) model, with $\Lambda$ referring to the Dark Energy,  is strongly supported  by the observations of  Cosmic Microwave Background (CMB) temperature and polarization anisotropies.
A primordial phase of rapid and exponential expansion, known as inflation~\citep{Guth:1980zm}, is currently associated with the mechanism producing primordial density fluctuations together with perturbations of the fabric of space-time, in the form of primordial gravitational waves. While these density fluctuations are associated with the curl-free component of the CMB polarization, $E$ modes, the gravitational waves left a divergence-less pattern and are referred to as $B$ modes~\citep{Zaldarriaga:1996xe}.\par
To date, $E$ modes have been characterized by several ground \citep{Louis_2017, Ade_2017,PhysRevD.101.122003,PhysRevLett.127.151301} and space-based \citep{Bennett_2013,Planck_Emodes} experiments. However, primordial $B$ modes have not yet been detected; only upper limits have been established on their amplitudes, with the tensor-to-scalar ratio, $r$ (the ratio of power in primordial gravitational waves to primordial density perturbations), constrained to $r < 0.032$ at 95\% confidence level by~\citep{Tristram:2021tvh}. %BICEP:2021xfz
Instrumental systematics and sensitivity together with polarized microwave emission from our own Galaxy pose a challenge for the detection of cosmological  $B$-modes.\par

The Lite (Light) satellite for the study of $B$-mode polarization and Inflation from cosmic background Radiation Detection (\lb) is a satellite mission that aims at detecting the CMB $B$ modes with a total error on the tensor-to-scalar ratio of $\delta r \sim0.001$~\citep{LiteBIRD:2022cnt}. It will observe the full sky in 15 frequency bands from 34 to 448~GHz for 3 years from the L2 Lagrangian point of the Sun-Earth system, with effective polarization sensitivity of $2.2\ \mathrm{\mu K-arcmin}$ and angular resolution of 31 arcmin (at 140~GHz), employing 4508 detectors sampling at 19.1~Hz~\citep{LiteBIRD:2022cnt}.\par
In order to achieve the $B$-mode requirements on $r$, the \lb\ mission requires unprecedented control of instrumental systematics of a typical CMB satellite.
Consequently, generating realistic \lb\ observation simulations that account for these systematic effects are crucial. These simulations are decisive for establishing stringent requirements on instrument parameters associated with systematic errors.\par

{To validate the design of its instruments, the \lb\ collaboration has developed a simulation pipeline. This computer program emulates the spacecraft's instrument operations, generating a synthetic data set that mirrors the anticipated \lb\ observations. Simulation pipelines are crucial for assessing the feasibility of achieving scientific objectives, optimizing observation strategies, and validating ground segment software and data storage systems. Developing a simulation pipeline involves defining scientific goals and instrument requirements, creating a mathematical model of the instrument, translating the mathematical model into a numerical model, developing and implementing the simulation pipeline software, and rigorously validating the accuracy and reliability of the pipeline.}

We thus used the \lb\ Simulation Framework (LBS\footnote{\url{https://github.com/litebird/litebird\_sim}}, \cite{tomasi2025})   and developed an end-to-end simulation pipeline, designed to generate 500 full sky simulated maps at \healpix\footnote{\url{http://healpix.sourceforge.net}}~\citep{Gorski:2004by,Zonca2019} resolution \nside = 512. {Given the rapid evolution of the LBS framework due to active community contributions, we note that these simulations utilized version 0.11.0. While the latest version is 0.13.0 at the time of writing, we have ensured backward compatibility with our end-to-end pipeline.}

Additionally, the pipeline produces a set of TODs using the \lb\ Instrument Model database (IMo)
% \footnote{\url{https://github.com/litebird/litebird\_imo/tree/v1.3}}
that is made available to \lb\ collaborators for several applications. In particular, we highlight that simulated TODs have been employed to assess the computational feasibility of an
end-to-end Bayesian analysis of the \lb\ experiment within the
Cosmoglobe framework \citep{2023A&A...678A.169E}. They will be presented in a companion paper  to be soon submitted \cite{Aurlien_2025}.
This paper is organised as follows. In Section~\ref{sec:instrument_model} we describe the instrument model adopted for this first release of simulations. In Section~\ref{sec:input_sky_maps} we present the beam convolved CMB and foreground maps used as input for the simulation pipeline. In Section~\ref{sec:simulation_pipeline} we illustrate the simulation pipeline. In Section~\ref{sec:results_and_validation} we show the outputs of the pipeline along with their validation. In Section~\ref{sec:conclusions} we draw conclusions. Appendix~\ref{sec:products} lists all products of the pipeline, Appendix~\ref{sec:maps} shows a comparison between the input and the scanned output maps.
\section{Instrument model}\label{sec:instrument_model}

For the simulations presented in this work, we employ the official \lb\ Instrument Model (IMo) database, which provides a detailed description of the current baseline instrument design (version \text{v1.3}).
\lb\ is organized into 3 separate telescopes: the Low Frequency Telescope (LFT), Medium Frequency Telescope (MFT), and High Frequency Telescope (HFT). The former employs reflective optics, whereas the latter are refractors.
%\sout{ In the IMo, detectors are initially categorized based on the telescope they are associated with: the Low Frequency Telescope (LFT), Medium Frequency Telescope (MFT), and High Frequency Telescope (HFT). Within each telescope, detectors are further organized into various collections based on the specific frequency they observe. These frequency-specific groupings are referred to as channels. Each channel is denoted by a letter indicating the telescope type (L for LFT, M for MFT, and H for HFT), a number indicating~\Marco{what exactly?}, and three additional numbers representing the channel's sensitivity frequencies.}
\par
\begin{figure}[h!]
    \centering
    $\vcenter{\hbox{\includegraphics[width=0.35\textwidth]{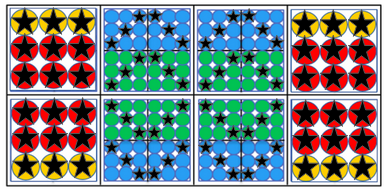}}}$
    $\vcenter{\hbox{\includegraphics[width=0.3\textwidth]{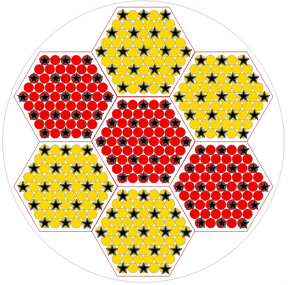}}}$
    $\vcenter{\hbox{\includegraphics[width=0.3\textwidth]{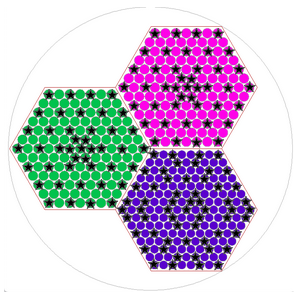}}}$
    \caption{Detectors present in \lb's LFT (left), MFT (middle) and HFT (right). The colors highlight wafers with the same frequency. The black stars indicate the selected pixels.}
    \label{fig:dets_configuration}
\end{figure}

As these simulations are mainly meant to be a pathfinder, we consider 1/3 of the nominal mission time (1 instead of 3 years) and of the detectors in the focal plane. In particular, we used all of the detectors for those channels with a maximum number
of 48, whereas for the rest we selected detectors in a configuration preserving the focal plane symmetry. Figure~\ref{fig:dets_configuration} shows the LFT, MFT and HFT focal planes (represented by colored circles), and we marked the camera pixels considered for the simulations with a black star. We remind the reader that each pixel in the focal plane is associated with 2 bolometer detectors.
\par
We report in the first column of Table~\ref{tab:channel_characteristics} the names of the 22 \lb\ channels. Table~\ref{tab:channel_characteristics} also summarises nominal specifics of each frequency channel employed for our simulations, i.e. the nominal number of detectors $\mathrm{n_{det}^{IMo}}$, the number of detectors employed in the end-to-end simulation $\mathrm{n_{det}^{e2e}}$, the beam Full Width Half Maximum (FWHM) in arcminutes, the detector Noise~Equivalent~ Temperature~(NET) values $\mathrm{NET^{IMo}}$ in $\mathrm{\mu K} \sqrt{s}$ and the detector NET actually employed in the simulations in $\mathrm{\mu K} \sqrt{s}$. The noise levels we use in the simulation, $\mathrm{NET^{e2e}}$, correspond to the IMo noise levels, $\mathrm{NET^{IMo}}$, rescaled to imitate the full focal plane 3-year mission as
\begin{equation}
    \text{noise rescaling} = \sqrt{\frac{1}{3\epsilon}\frac{\mathrm{n_{det}^{e2e}} }{ \mathrm{n_{det}^{IMo}}} },
\end{equation}
where the $1/3$ takes into account the reduced simulated time, $\mathrm{n_{det}^{e2e}}/\mathrm{n_{det}^{IMo}}$ the reduced number of simulated detectors and $\epsilon$ is the detector efficiency. Given the goals of this round of simulations, we do not include any  optical efficiency and assume an ideal detector yield.
\begin{table}[!ht]
    \centering
    \renewcommand{\arraystretch}{1.5} %increase the cell height
    \begin{tabular}{c|ccccc}\hline
       Channel & $\mathrm{n_{det}^{IMo}}$ & $\mathrm{n_{det}^{e2e}}$ & Beam & $\mathrm{NET^{IMo}}$ & $\mathrm{NET^{e2e}}$ \\
       &&&[arcmin]&$\mathrm{[\mu K \sqrt{s}]}$&$\mathrm{[\mu K \sqrt{s}]}$\\\hline
       L1-040 &  48 &  48 & 70.5 & 114.6 & 66.2 \\
       L1-060 &  48  & 48 & 51.1 & 65.3 & 37.7 \\
       L1-078 &  48  & 48 & 43.8 & 58.6 & 33.8 \\
       L2-050 &  24  & 24 & 58.5 & 72.5 & 41.9 \\
       L2-068 &  24  & 24 & 47.1 & 68.8 & 39.7 \\
       L2-089 &  24  & 24 & 41.5 & 62.3 & 36.0 \\
       L3-068 &  144 & 48 & 41.6 & 105.6 & 35.2 \\
       L3-089 &  144 & 48 & 33.0 & 65.2 & 21.7 \\
       L3-119 &  144 & 48 & 26.3 & 40.8 & 13.6 \\
       L4-078 &  144 & 48 & 36.9 & 82.5 & 27.5 \\
       L4-100 &  144 & 48 & 30.2 & 54.9 & 18.3 \\
       L4-140 &  144 & 48 & 23.7 & 38.4 & 12.8 \\\hline
       M1-100 & 366 & 126 & 37.8 & 71.7 & 24.3 \\
       M1-140 & 366 & 126 & 30.8 & 54.0 & 18.3 \\
       M1-195 & 366 & 126 & 28.0 & 59.6 & 20.2 \\
       M2-119 & 488 & 168 & 33.6 & 55.7 & 18.9 \\
       M2-166 & 488 & 168 & 28.9 & 54.4 & 18.4 \\\hline
       H1-195 & 254 &  86 & 28.6 & 74.0 & 24.9 \\
       H1-280 & 254 &  86 & 22.5 & 97.3 & 32.7 \\
       H2-235 & 254 &  86 & 24.7 & 76.1 & 25.6 \\
       H2-337 & 254 &  86 & 20.9 & 154.6 & 52.0 \\
       H3-402 & 338 & 116 & 17.9 & 385.6 & 130.5 \\\hline

       % L1-040 &  48 &  48 & 70.5 & 114.63 & 66.18 \\
       % L1-060 &  48  & 48 & 51.1 & 65.28 & 37.69 \\
       % L1-078 &  48  & 48 & 43.8 & 58.61 & 33.84 \\
       % L2-050 &  24  & 24 & 58.5 & 72.48 & 41.85 \\
       % L2-068 &  24  & 24 & 47.1 & 68.81 & 39.73 \\
       % L2-089 &  24  & 24 & 41.5 & 62.33 & 35.99 \\
       % L3-068 &  144 & 48 & 41.6 & 105.64 & 35.21 \\
       % L3-089 &  144 & 48 & 33.0 & 65.18 & 21.73 \\
       % L3-119 &  144 & 48 & 26.3 & 40.78 & 13.59 \\
       % L4-078 &  144 & 48 & 36.9 & 82.51 & 27.5 \\
       % L4-100 &  144 & 48 & 30.2 & 54.88 & 18.29 \\
       % L4-140 &  144 & 48 & 23.7 & 38.44 & 12.81 \\\hline
       % M1-100 & 366 & 126 & 37.8 & 71.7 & 24.29 \\
       % M1-140 & 366 & 126 & 30.8 & 54.0 & 18.29 \\
       % M1-195 & 366 & 126 & 28.0 & 59.61 & 20.19 \\
       % M2-119 & 488 & 168 & 33.6 & 55.65 & 18.85 \\
       % M2-166 & 488 & 168 & 28.9 & 54.37 & 18.42 \\\hline
       % H1-195 & 254 &  86 & 28.6 & 73.96 & 24.85 \\
       % H1-280 & 254 &  86 & 22.5 & 97.26 & 32.67 \\
       % H2-235 & 254 &  86 & 24.7 & 76.06 & 25.55 \\
       % H2-337 & 254 &  86 & 20.9 & 154.64 & 51.95 \\
       % H3-402 & 338 & 116 & 17.9 & 385.6 & 130.45 \\\hline
    \end{tabular}
    \caption{Channel characteristics. The nomenclature in the first column expresses the telescope identifier (first letter for Low, Medium or High Frequency Telescope), wafer number (first digit), and central frequency (in GHz) each channel is sensitive to. For each channel, $\mathrm{n_{det}^{IMo}}$ is the number of detectors reported in the IMo (totaling 4508), $\mathrm{n_{det}^{e2e}}$ is the number of detectors employed for our simulations (totaling 1678), Beam is the beam FWHM in arcminutes, $\mathrm{NET^{IMo}}$ is the noise requirement from the IMo, $\mathrm{NET^{e2e}}$ is the noise baseline used for our simulations (in units of $\mathrm{\mu K \sqrt{s}}$).}
    \label{tab:channel_characteristics}
\end{table}\par
\begin{figure}[htbp]
    \centering
    \includegraphics[width=0.6\textwidth]{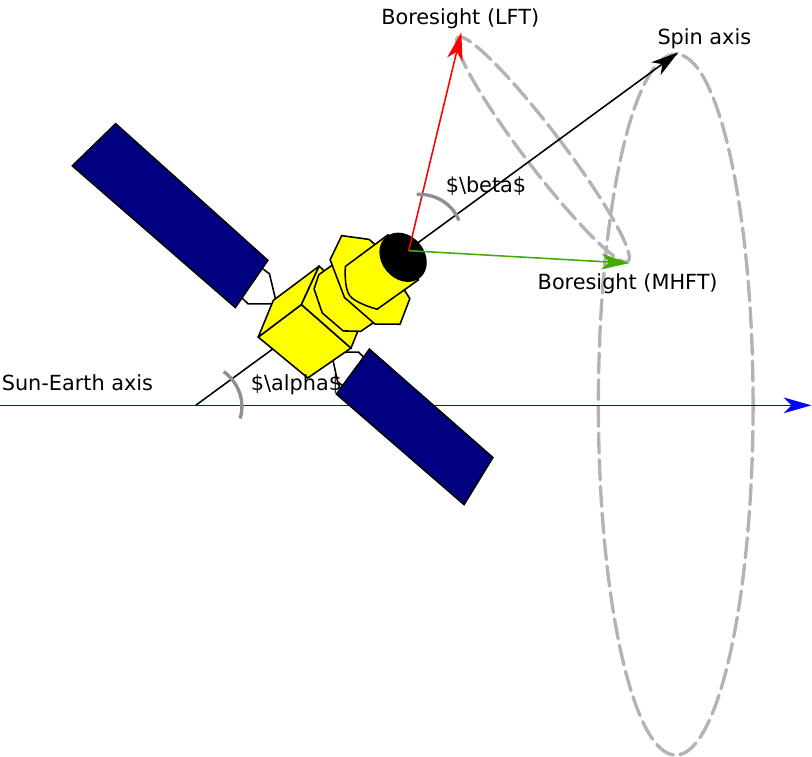}
    \caption{Cartoon representation of the \lb\ satellite and its scanning parameters. $\alpha=45^\circ$ is the angle between the spin axis and the Sun--Earth axis, $\beta=50^\circ$ is the angle separating the boresight from the spin axis. A summary of the scanning parameters is in Table~\ref{tab:scanning_parameters}.}
    \label{fig:scanning strat}
\end{figure}

For the scanning strategy parameters in these simulations we consistently employ the ones presented in~\cite{LiteBIRD:2022cnt} and summarized in Table~\ref{tab:scanning_parameters} and Figure~\ref{fig:scanning strat}. The telescope boresight is positioned at an angle of $\beta = 50^\circ$ relative to the spin-axis, rotating at a rate of $0.05$ rpm (20 min.). The spin-axis undergoes precession around the Sun--Earth direction with an angle of $\alpha = 45^\circ$, completing one full rotation in $\sim 3.2$ hours. Combining three motions (spin-axis rotation, precession around the Sun--Earth axis and 1-year revolution around the Sun) enables the boresight to cover the entire sky within six months. For further details about the \lb\ scanning strategy refer to~\citep{LiteBIRD:2024fzn}.

\begin{table}[htbp]
\centering
\begin{tabular}{cccccccc}
\hline
$\alpha$ & $\beta$ & Precession period & Spin rate & Sampling rate & \multicolumn{3}{c|}{HWP rotation rate [rpm]} \\
\cline{6-8}
[deg.] & [deg.] & [min.] & [rpm] & [Hz] & LFT & MFT & HFT \\
\hline
45 & 50 & 192.348 & 0.05  & 19.1 & 46 & 39 & 61 \\
\hline
\end{tabular}
\caption{Parameters of the observation strategy and sampling rate of the \lb\ satellite.}
\label{tab:scanning_parameters}
\end{table}\par

\section{Input sky maps}\label{sec:input_sky_maps}
Our simulation pipeline takes as input a set of $(I,Q,U)$ sky maps at \healpix~\citep{Gorski:2004by,Zonca2019} resolution \nside = 512 for each of the 22 \lb\ frequency channels. We generate the input maps of CMB, Galactic and extragalactic foregrounds separately. The different tools we use to simulate the different components are presented below.\par
We employ the Map-Based Simulation (MBS) module within the LBS framework to generate maps of Galactic diffuse foregrounds. MBS serves as a wrapper around the Python Sky Model (PySM)\footnote{www.github.com/galsci/pysm} package~\citep{pysm3,Zonca_2021,Thorne_2017}. We consider the following models:
\begin{description}
\item[\texttt{d1}] thermal dust modelled as a single-component modified black-body with varying temperature and spectral index across the sky, based on the maps from the Planck 2015 analysis~\citep{planck2014-a12};
\item[\texttt{s1}] synchrotron modelled as a power law model with varying spectral index and no curvature, based on Haslam 408 MHz \citep{remazeilles2015} and WMAP data~\citep{delabrouille_2013, miville_2008};
\item[\texttt{a1}] two unpolarised spinning dust populations of Anomalous Microwave Emission (AME), based on the templates obtained from \emph{Planck} with  Commander methodology~\citep{planck2014-a12};
\item[\texttt{f1}] unpolarised free-free emission with a constant spectral index of $-2.14$, based on the templates obtained from \emph{Planck} with  Commander methodology~\citep{planck2014-a12};
\item[\texttt{co1}] $J:1 \rightarrow
0, \, 2\rightarrow1, \, 3\rightarrow2$ rotational lines of Galactic CO emission, whose center frequency is $115.3$, $230.5$ and $345.8$ GHz, respectively~\citep{puglisi_2017, planck2013-p03a}.
\end{description}\par
We use the WebSky simulations~\citep{Stein_2020} and PySM to generate the maps of extragalactic emissions across the \lb\ frequency band. We produce maps of thermal (tSZ) and kinetic Sunyaev–Zeldovich (kSZ) emission, Cosmic Infrared Background (CIB) and lensing convergence by populating  Dark Matter haloes from an N-body simulation (for further details please refer to \cite{Stein_2020}). The tSZ effect arises from the inverse Compton scattering of CMB photons off hot electrons in galaxy clusters, distorting the CMB spectrum~\cite{Birkinshaw:1998qp}. The kSZ effect, on the other hand, is caused by the Doppler shift of CMB photons due to the motion of these same galaxy clusters relative to the CMB rest frame~\cite{Sunyaev:1980vz}. The CIB is produced by the cumulative emission of dust-covered star-forming galaxies across cosmic time~\cite{Lagache:2005sw}. The lensing convergence map represents the projected mass density along the line of sight~\cite{Lewis:2006fu}, decisive for modeling and correcting the distortions in the CMB caused by gravitational lensing, allowing for a more accurate reconstruction of the matter distribution and extraction of cosmological information.
We also use recent radio sources catalogs to produce frequency-dependent maps~\citep{Li_2022,puglisi_2018,Lagache:2019xto}.\par
\begin{figure}[!ht]
    \centering
    \includegraphics[width=1.\textwidth]{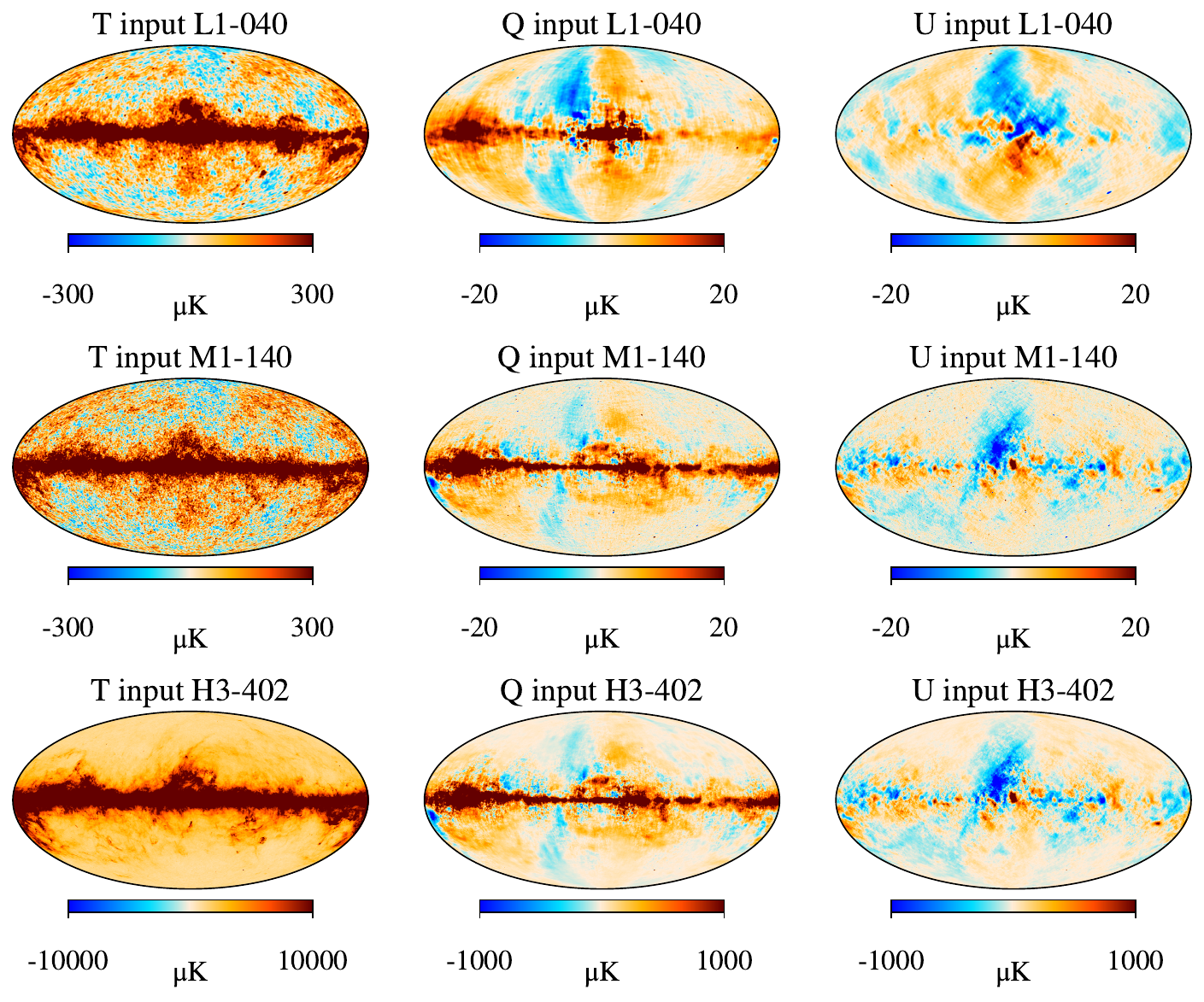}
    \caption{Coadded input maps in Galactic coordinates. $T$ (left column), $Q$ (middle column), $U$ (right column) coadded input maps of CMB and foregrounds for the first simulation. First row: L1-040 channel of LFT. Second row: M1-140 channel of MFT. Third row: H3-402 channel of HFT.
    }
    \label{fig:maps_input_coadded}
\end{figure}

We produce 500 Monte Carlo (MC) realisations of the CMB sky from the \Planck\ 2018 cosmological parameters~\cite{planck2016-l06}, lensed with the same WebSky convergence map via the \texttt{lenspyx}\footnote{https://github.com/carronj/lenspyx} 
 package~\citep{Reinecke:2023gtp}. By using the beam specifications provided by the IMo and reported in Table~\ref{tab:channel_characteristics}, we convolve the input CMB and foreground maps. Then, by scanning the
maps, we produce TOD that already include the response to
instrumental beams. The beams are assumed to be circular Gaussian with the same FWHM for all detectors in a given channel. We show in Figure~\ref{fig:maps_input_coadded} the coadded input maps of CMB and foregrounds for three different channels chosen from the three telescopes.

\section{Simulation pipeline}\label{sec:simulation_pipeline}
The pipeline developed for the production of the simulations heavily relies on the LBS, integrated with the IMo described in Section~\ref{sec:instrument_model}. The initial step of the pipeline involves loading the instrument parameters summarized in Tables~\ref{tab:channel_characteristics},~\ref{tab:scanning_parameters}, including the quaternions describing the rotation from the Ecliptic reference frame to the reference frame of each detector, the rotation frequency of the HWPs\footnote{The polarized incident radiation is modulated at four times the rotation frequency of the HWP, effectively shifting the polarized sky signal to a narrow frequency band, specifically above the $1/f$ noise knee frequency. This modulation significantly improves the measurement of CMB polarization by separating the sky signal from the $1/f$ noise.} (46/39/61 rpm for LFT/MFT/HFT) and the detector noise levels.\par
These quantities are firstly used to calculate the pointing information for each simulated detector and sample, employing the \lb\ scanning strategy (see~\cite{LiteBIRD:2022cnt} for further details) and the Ephemeridis tables to get an accurate position of the spacecraft around the Sun at each instant. In particular, we calculate the colatitude $\theta$ and longitude $\phi$ as well as the polarization angle $\psi$.\par
Subsequently, the beam convolved input maps presented in Section~\ref{sec:input_sky_maps} are scanned (see Appendix~\ref{sec:maps} and Figure~\ref{fig:maps_input_vs_output}) using the pointing information to generate TOD for CMB (\texttt{tod\_cmb}) and foregrounds (\texttt{tod\_fg}). Additionally, three noise timelines are simulated using the detector noise characteristics $\mathrm{NET^{e2e}}$ reported in Table~\ref{tab:channel_characteristics}: \texttt{tod\_wn}, which consists of white noise only; \texttt{tod\_wn\_1f\_30mHz} and \texttt{tod\_wn\_1f\_100mHz}, containing white noise and correlated $1/f$ noise with a knee frequency $f_{\rm knee}$ of 30 mHz and 100 mHz, respectively. The \texttt{tod\_wn\_1f\_30mHz} timeline represents a realistic scenario, with a knee frequency close to what is expected from \lb. The \texttt{tod\_wn\_1f\_100mHz} timeline is instead a pessimistic case for which the $1/f$ noise is worse and it is useful for setting requirements in pessimistic conditions. A summary of all products may be found in the Appendix~\ref{sec:products}.

We used FFTs to generate the $1/f$ noise, leveraging a simulation length equal to the length of the TOD, which corresponds to one year of data at a sampling rate of 19.1 Hz. To handle the inherent limitation of FFTs, which restrict the simulation to the length of the TOD, we selected a simulation length equal to the TOD length (one year) and padded it to the nearest power of 2 to optimize FFT performance. By ensuring that the FFT length matches the data length, we avoid issues with periodicity. The decision to use FFTs for generating $1/f$ noise was driven by its computational efficiency, especially when compared to the alternative approach of Markov random walks, which, although more flexible for very low frequencies, would have been significantly more computationally demanding. In frequency domain the $1/f$ noise is:
\begin{equation} \label{eq:1/f noise}
    \tilde{d}_i \rightarrow \tilde{d}_i \times \sigma \sqrt{\frac{f_i^\alpha+f_{\mathrm{knee }}^\alpha}{f_i^\alpha+f_{\mathrm{min }}^\alpha}} \quad \mathrm{ for } \quad i>0, \quad \mathrm{ with } \quad \tilde{d}_0=0
\end{equation}
where $\tilde{d}_i$ is the Fourier transform of the TOD at frequency index $i$, $f_i$ is the frequency, $f_\mathrm{knee}$ is the knee frequency where the $1/f$ noise power equals the white noise power in the power spectrum density (see Figure~\ref{fig:tod_noise_frequency_space}), $f_\mathrm{min}$ is the the low-frequency cutoff, $\sigma$ is the rescaled white noise level ($\mathrm{NET}^\mathrm{e2e}$ in Table~\ref{tab:channel_characteristics}) and $\alpha=1$ is the low-frequency spectral tilt.

For the first simulation only, we produce a timestream \texttt{tod\_dip} with the CMB dipole signal. We simulate the dipole signal using the \texttt{TOTAL\_FROM\_LIN\_T} model\footnote{https://litebird-sim.readthedocs.io/en/master/dipole.html}, following the LBS nomenclature. This is described by:
\begin{equation}
    \Delta T = \dfrac{T_0}{f(x)}\left(\dfrac{B(\nu \gamma(1- \vec{\beta} \cdot \hat{n}),T_0)}{(\gamma(1- \vec{\beta} \cdot \hat{n}))^3 BB(\nu,T_0)}\right),
\end{equation}
where $\Delta T$ is the dipole signal, $T_0$ is the CMB temperature \cite{fixsen2009}, $\vec{\beta}=\vec{v} / c$ is the velocity vector (relative to the speed of light $c$) accounting for Earth's motion in the CMB rest frame, $\gamma=(1-\vec{\beta} \cdot \vec{\beta})^{-1/2}$, $\hat{n}$ is the line-of-sight direction, $\nu$ is the CMB frequency, $x=h \nu / k_B T$ so that:
\begin{equation}
    f(x)=\frac{x e^x}{e^x-1}
\end{equation}
where $h$ is the Planck constant and $k_B$ is the Boltzmann constant, and the black-body spectrum $B$ is given by:
\begin{equation}
    B(\nu, T)=\frac{2 h \nu^3}{c^2} \frac{1}{e^{h \nu / k_B T}-1}=\frac{2 h \nu^3}{c^2} \frac{1}{e^x-1}.
\end{equation}\par
\begin{figure}[h]
    \centering
    \includegraphics[width=1.0\linewidth]{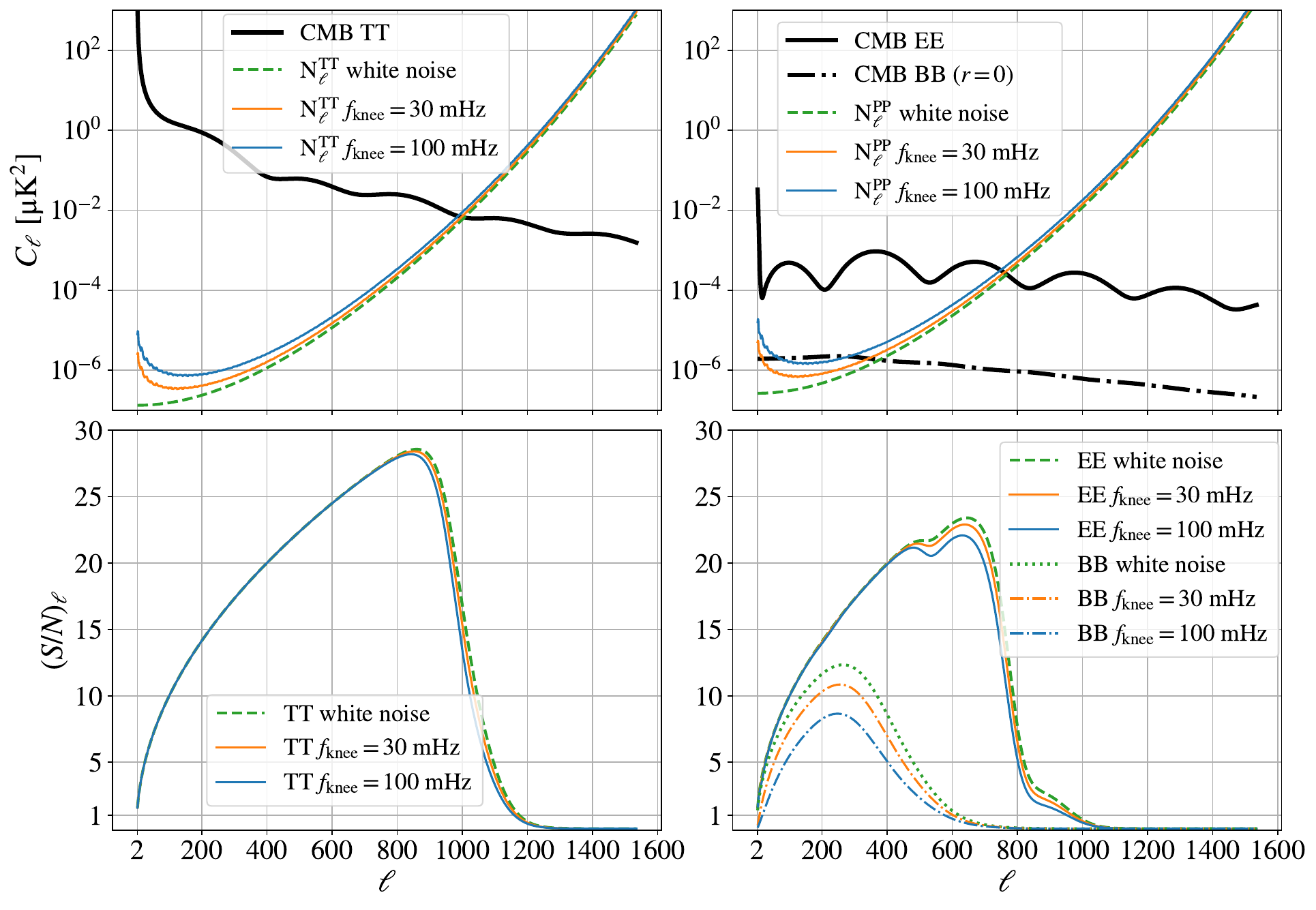}
    \caption{Top panels: the CMB power spectra for temperature and polarization are compared to the noise spectra in the absence of a half-wave plate (HWP). The fiducial signal is derived from the best-fit cosmological parameters of the \Planck\ 2018 data~\cite{planck2016-l06}, obtained from the TT, TE, EE + lowE + lensing analysis. The noise power spectra are computed as the spectrum of the difference between the binned output and input maps, inversely coadded across all channels and averaged over the first 50 simulations. The various curves represent different levels of $1/f$ noise knee frequencies, along with the white noise-only baseline. For polarization (without HWP), the noise is estimated by scaling the temperature noise by a factor of 2. This scaling is consistent with the elevated noise levels observed in polarization, illustrated in Figure~\ref{fig:covs} and by the green curves in Figure~\ref{fig:covs_validation_test}.\\
    Bottom panels: \lb\ high signal-to-noise ratio is evident across all $1/f$ noise levels. Cosmic variance dominates the very low-$\ell$ regime, and the signal remains robust up to multipoles of approximately $\ell \simeq 1100$ (TT), $\ell \simeq 900$ (EE), and $\ell \simeq 500$ (BB).}
    \label{fig:litebird_SNR}
\end{figure}

A binner map-making algorithm\footnote{A binner (or naive) map-maker constructs sky maps from TODs by averaging observed data points into pixels on a sky map. This approach does not incorporate specific noise modeling (e.g., assuming only white noise) or deconvolve instrument effects. While computationally efficient, it is less accurate compared to advanced techniques such as maximum-likelihood or optimal map-makers.} is applied to the TOD to produce maps containing CMB, foregrounds, white noise, and $1/f$ noise at a \healpix~\citep{Gorski:2004by,Zonca2019} resolution of \nside = 512. Although the binner map-maker inherently assumes uncorrelated noise in the timelines, the presence of the HWP in the polarization case is expected to effectively suppress the $1/f$ noise component. For temperature, the $1/f$ noise becomes negligible due to the extremely favorable signal-to-noise ratio (SNR), as illustrated in Figure~\ref{fig:litebird_SNR}. The top left plot compares the CMB temperature with the noise power spectra. The bottom left panel depicts the SNRs of both levels of $1/f$ noise with respect to the white noise baseline. The SNR is computed as:
\begin{equation}
\begin{split}
    & \text{SNR} = \frac{C_{\ell}}{\sigma}\\
    & \sigma^2 = \frac{2}{2\ell+1}(C_{\ell} + \mathrm{N}_{\ell})^2.
\end{split}
\end{equation}
The noise power spectra in both panels for the $1/f$ noise levels are obtained by inversely coadding the spectra of the binned output and input map differences across all channels, and averaged over 50 simulations:
\begin{equation}
    C_\ell = \left[ \sum_\mathrm{chann} \frac{1}{C_{\ell,\, \mathrm{chann}}} \right]^{-1}.
\end{equation}
As a test case, we consider a scenario analogous to the temperature case but without the presence of a HWP. The polarization noise is derived by doubling the power of the temperature noise. This approach is consistent with the observed increase in noise levels for polarization, as demonstrated in Figure~\ref{fig:covs} and by the green curves in Figure~\ref{fig:covs_validation_test}.\par

We acknowledge that, at the time of submission, a bug was identified in the noise time-stream generation code. This resulted in the unintentional simulation of a higher level of $1/f$ noise, mainly  affecting the temperature simulations. However, as evidenced by the unprecedented \lb\ signal-to-noise ratios presented in Figure~\ref{fig:litebird_SNR}, the impact of this error is considered minor for both polarization and temperature data. 

The binner is utilized for all the simulations, in particular producing 500 maps for each of the two cases,  with $f_{\rm knee} = 30$ mHz and  with $f_{\rm knee} = 100$ mHz. Furthermore, only for the first simulation, we stored the TODs, pointing information and noise covariance matrices. In detail, we generated a noise covariance matrix (in pixel space) for each channel, which is crucial for realistic map-based simulations that incorporate pixel-to-pixel correlations in the $I,\, Q$, and $U$ maps. Map-based simulations are indispensable for testing and validating the observational system's performance, assessing systematic effects, and ensuring the robustness of data analysis pipelines.

Each result presented in this work is fully reproducible, provided that the noise characteristics outlined in Table~\ref{tab:channel_characteristics} are adhered to. 
The repository employing the simulation script and the configuration file has been made publicly available\footnote{\texttt{\color{blue}{github.com/litebird/e2e-simulation}}} .
  The public version\footnote{\url{github.com/litebird/instrumentdb}} of the IMo is implemented in the LBS, allowing for easy replication of the noise properties.
 However, while the noise features are reproducible, variations in the focal plane geometry are not fully captured in the IMO version of the code.
\section{Results and validation}\label{sec:results_and_validation}
We present here the results obtained with the simulation pipeline along with their validation. In Section~\ref{sec:tod} we show results related to TODs, e.g. their plot for different detectors or components. In Appendix~\ref{sec:maps}, we present the output maps and their difference with the input ones. In Section~\ref{sec:covariance_matrices}, we show the noise covariance matrices. In Section~\ref{sec:power_spectra}, we discuss the power spectra calculated from the maps. Finally, in Section~\ref{sec:computational_cost} we report the computational cost.

%\peppe{no need of following lines ... }
%\sout{All the results saved to disk are in units of K. However, we report them here in units of $\mu$K for convenience.}\sout{We do not show the results for all the channels, detectors, samples and simulations that we produced because they are similar to the ones reported in the following.} \sout{Moreover, the maps are available at \texttt{NERSC}\footnote{For details consult the Wiki page:} \url{https://wiki.kek.jp/display/litebird/Post\-PTEP++Simulations}.
%\Marco{the minus sign after "Post" gives the error, I tried passing the hyphens option to url but it doesn't work. Also, }}at \url{/global/cfs/cdirs/litebird/simulations/LB_e2e_simulations/e2e_ns512}, and the plots can be obtained with the validation notebook present at \Marco{notebook validation GitHub? It's private...} 
%\Nicolo{If we make it public, we should probably merge the 2 branches..}.\Marco{what about the tests we did with noise only maps? Or with cmb+fg+wn?} \Nicolo{I think in the validation section anyway} \Marco{the question was "do we put them or not? we have the cmb+fg+wn maps only from few sims, maybe 1}} \peppe{maps are not publicly  available so there ain't no problem .  

\subsection{TOD}\label{sec:tod}

\begin{figure}[!ht]
    \centering
    \includegraphics[width=1.\textwidth]{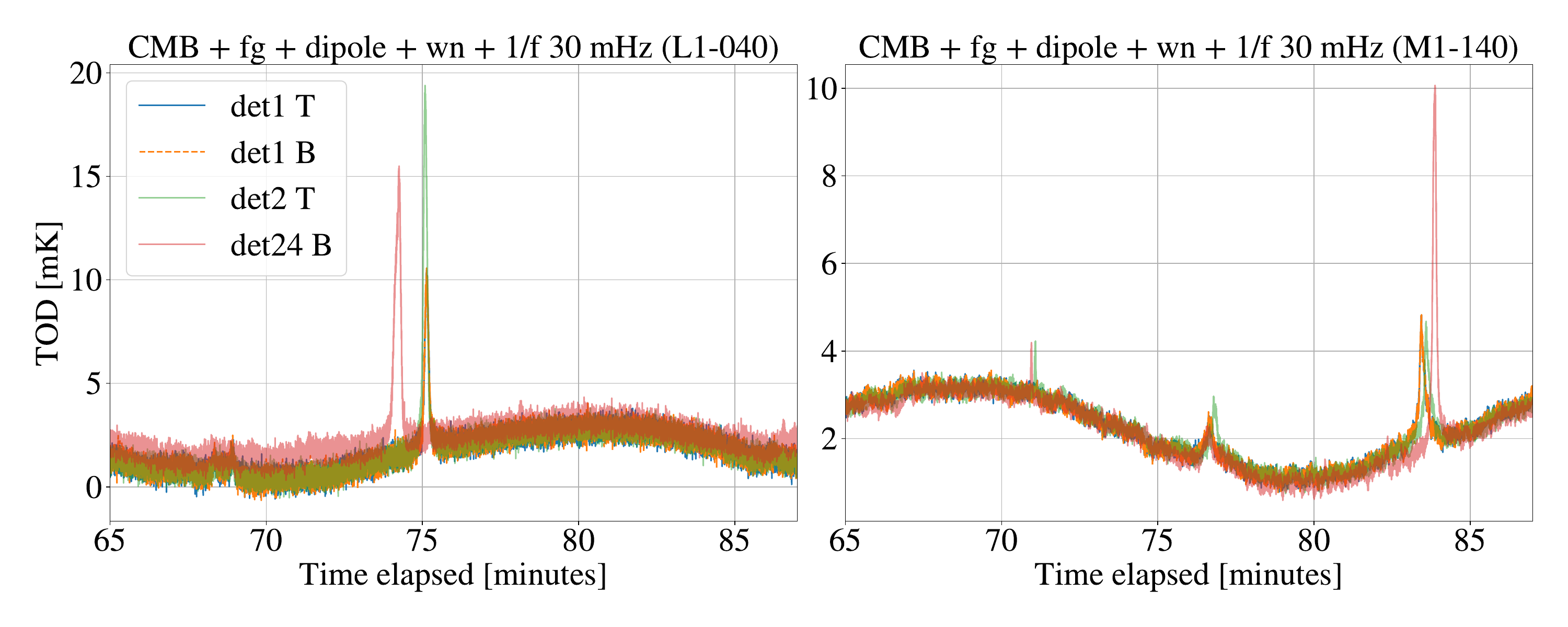}
    \vspace{-4mm}
    % \includegraphics[width=0.85\textwidth]{fig/tod_cmb_fg_dip_wn_1f30_different_det_M1-140.pdf}
    % \vspace{-4mm}
    % \includegraphics[width=0.85\textwidth]{fig/tod_cmb_fg_dip_wn_1f30_different_det_H3-402.pdf}
    \caption{Sum of TODs of CMB, foregrounds, dipole, white noise and $1/f$ noise with $f_{\rm knee}=30\ \mathrm{mHz}$ for three different detectors and for the first simulation. The numbers in the legend refer to the detectors number as explained in the text, while the T (B) letter refer to the detector being top (bottom). Left panel: L1-040 channel of LFT. Right panel: M1-140 channel of MFT.}
    \label{fig:tod_sum_different_det}
\end{figure}

\begin{figure}[!ht]
    \centering
    \includegraphics[width=1.\textwidth]{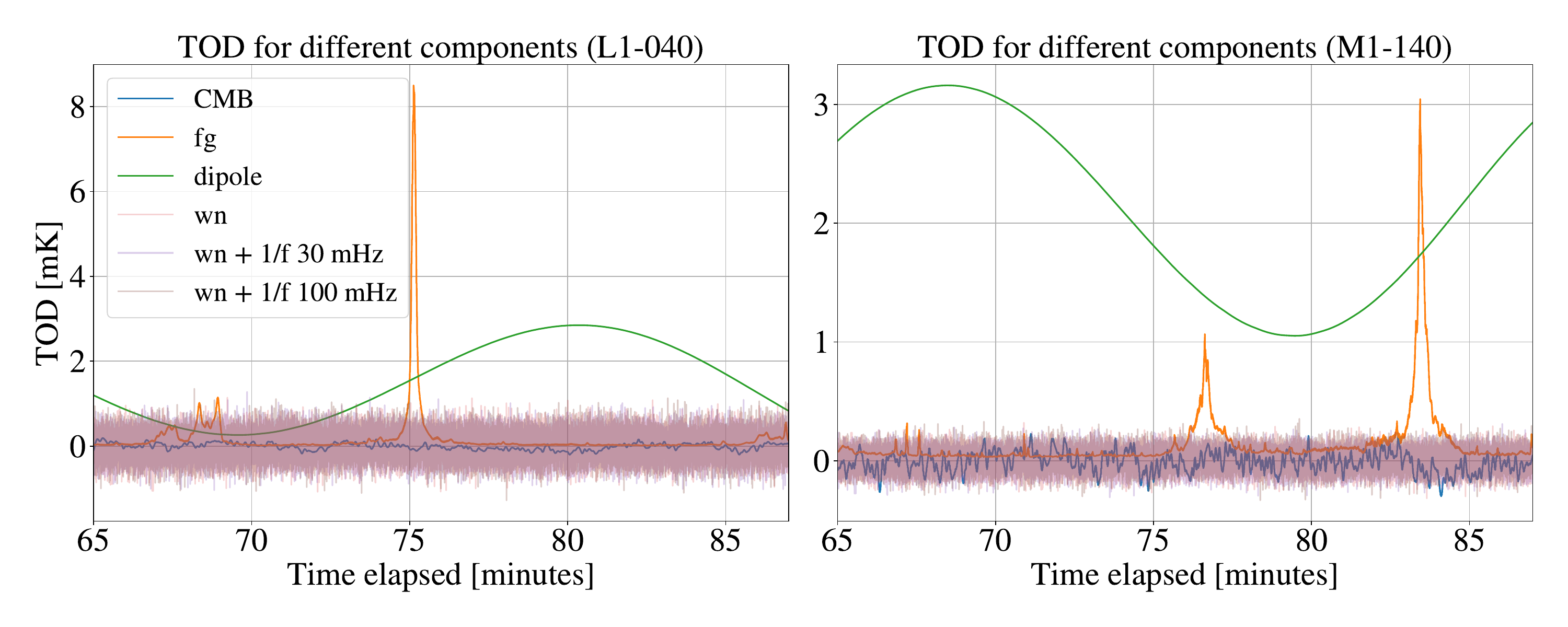}
    \vspace{-4mm}
    \caption{TOD of CMB, foregrounds, dipole, white noise, white noise + $1/f$ noise with $f_{\rm knee}=30\ \mathrm{mHz}$ and white noise + $1/f$ noise with $f_{\rm knee}=100\ \mathrm{mHz}$ for one different detector and for the first simulation. Left panel: L1-040 channel of LFT. Right panel: M1-140 channel of MFT.}
    \label{fig:tod_different_components}
\end{figure}

%\Nicolo{Should we say that the tods are stored at the oslo HPC?}\Marco{maybe yes, let's hear from the others}\peppe{no we can say that there is a paper to be submitted devoted to TOD analysis.}\\
We show in Figure~\ref{fig:tod_sum_different_det} the sum of the TODs of CMB, foregrounds, dipole, white noise and $1/f$ noise with $f_{\rm knee}=30$ mHz for three different detectors of LFT and MFT and for the first simulation, i.e. the only one for which we saved TODs to disk.
{We remind the reader that these TODs were also employed for assessing feasibility  of Cosmoglobe framework when applied onto \lb\ data and results will be presented in \cite{Aurlien_2025}}.
The numbers in the legend are related to the selected detectors in Figure~\ref{fig:dets_configuration}: the counting starts from 1 for the top-left detector marked with a black star and increases going to the right and going to the following row. The letters T or B denote the bolometer pair, identifying each detector as either \emph{top} or \emph{bottom}. For Q-type antennas, the top/bottom designation corresponds to orientations of $0^\circ/90^\circ$, while for U-type antennas, it indicates orientations of $45^\circ/135^\circ$. Detectors in the same pair (top and bottom) observe the same signal since they are aligned in the same direction, differing only in their noise component, even though is difficult to distinguish visually. The detectors shown in Figure~\ref{fig:tod_sum_different_det} are selected to illustrate this behavior.
When examining a single detector's time stream, various structures become evident. The long time scale modulation observed in Figure~\ref{fig:tod_sum_different_det} corresponds to the dipole signal, while the spikes are related to the samples obtained while scanning across the Galactic midplane (e.g., notice the L1-040 channel is strongly dominated by the Synchrotron signal, as the spikes are more prominent than the rest), small scale fluctuations are instead related to CMB and noise. Furthermore, the difference between the LFT and MFT panels, where the signals appear out of phase, is due to the instruments being oriented in opposite directions (recall Figure~\ref{fig:scanning strat}).
We show in Figure~\ref{fig:tod_different_components} the TOD of all the components separately, as described in Section~\ref{sec:simulation_pipeline} and for a single detector of LFT and MFT.\par
\begin{figure}[!ht]
    \centering
    \includegraphics[width=1.\textwidth]{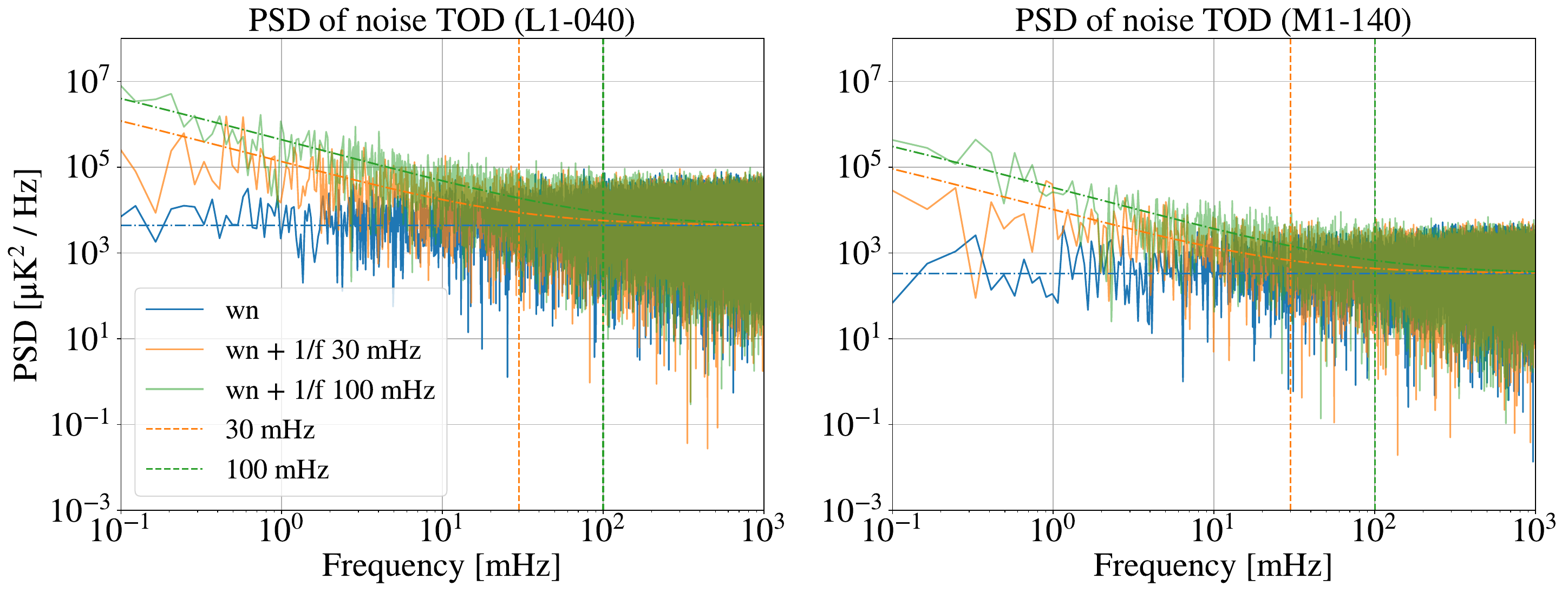}
    \caption{Power Spectral Density of the noise TOD shown in Figure~\ref{fig:tod_different_components}. The solid lines have been obtained from data, while the dashed-dotted lines show the analytical model (Eq.~\ref{eq:1/f noise}). The vertical dashed lines show the two knee frequencies used for the two $1/f$ noise timelines. Left panel: L1-040 channel of LFT. Right panel: M1-140 channel of MFT.}
    \label{fig:tod_noise_frequency_space}
\end{figure}

In Figure~\ref{fig:tod_noise_frequency_space} we show the Power Spectral Density (PSD) of the TOD noise shown in Figure~\ref{fig:tod_different_components}.
As expected, the white noise curve remains nearly constant across the entire frequency range. In contrast, the presence of the $1/f$ noise component causes an increase in noise power at lower frequencies, with a more pronounced effect for $f_{\rm knee}=100$ mHz compared to $f_{\rm knee}=30$ mHz. The plotted PSD reflects the expected behavior, as shown by the vertical and tilted lines corresponding to the two different knee frequencies of the $1/f$ noise.

\subsection{Covariance matrices}\label{sec:covariance_matrices}
\begin{figure}[!ht]
    \centering
    \includegraphics[width=1.0\textwidth]{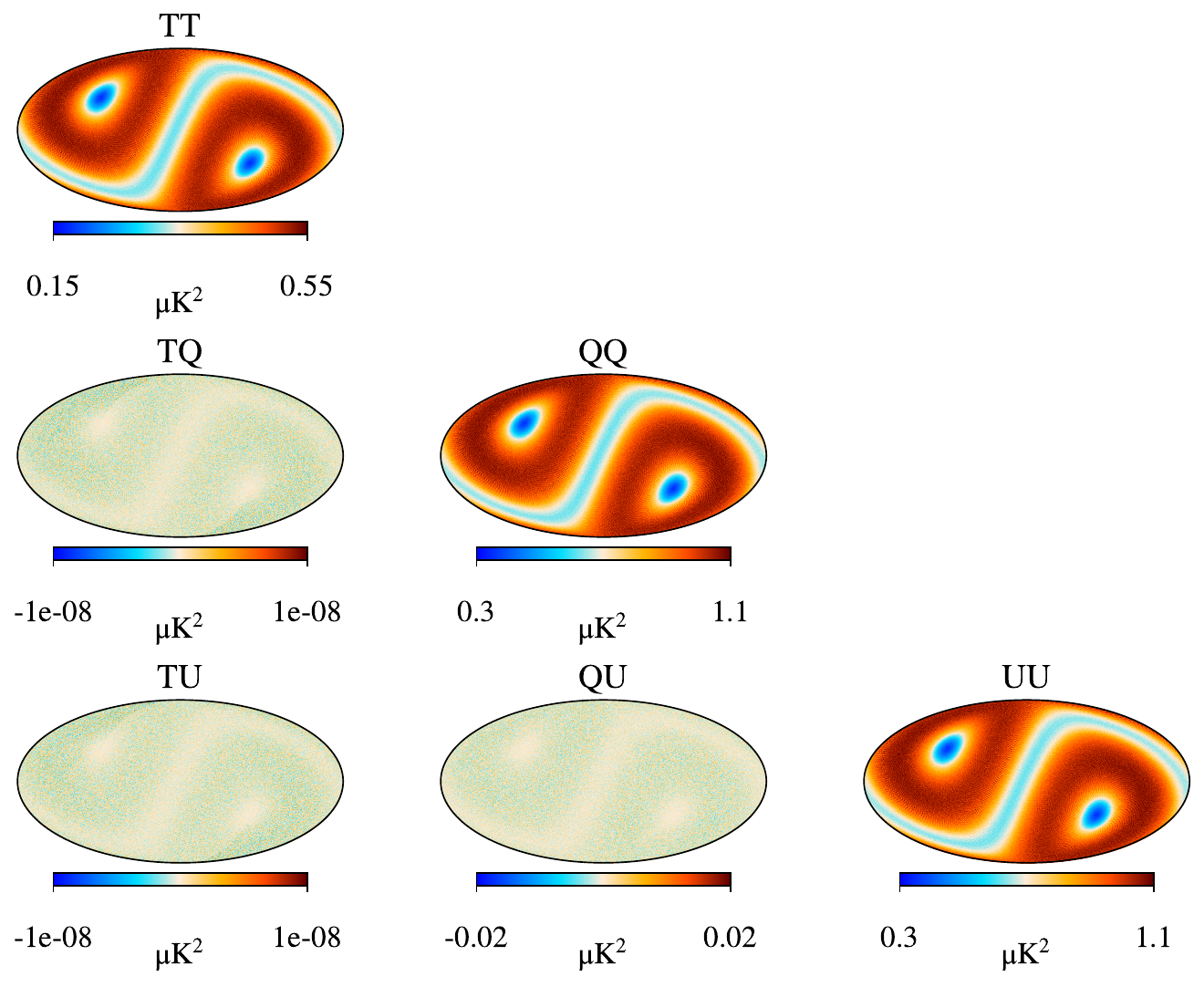}
    \caption{Noise covariance matrices in pixel-space in Galactic coordinates for the channel M1-140 of MFT. Note that the value ranges differ between the maps.}
    \label{fig:covs}
\end{figure}

We present the inverse of the noise covariance matrix, $\mathcal{N}_{pp}^{-1}$, in Figure~\ref{fig:covs}, specifically for the MFT M1-140 channel. This matrix is constructed from simulated pixel correlations, where each pixel's covariance is a $3\times3$ matrix.
This is readily observable, beginning with mapmaking equations:
\begin{equation} \label{eq: mapmaking}
    \hat{\mathbf{m}} = (\mathbf{A}^\mathrm{T}\mathbf{N}^{-1} \mathbf{A})^{-1} \: \mathbf{A}^\mathrm{T} \mathbf{N}^{-1}\mathrm{\mathbf{d}},
\end{equation}
where $\hat{\mathbf{m}}$ is the estimated map, $\mathbf{A}$ is the pointing matrix and contains information about how each pixel in the sky contributes to the time-ordered signal $\mathrm{\mathbf{d}}$ collected by one detector, $\mathbf{N}= \mathbf{N}_{tt}=\langle \mathbf{n\,n}^\mathrm  T\rangle$ is the noise covariance matrix in time domain and $\mathcal{N}_{pp}^{-1}$ $=(\mathbf{A}^\mathrm{T}\mathbf{N}_{tt}^{-1} \mathbf{A})^{-1}$  noise covariance matrix in pixel space.
 This block encompasses the auto-correlations ($TT,\, QQ,\, UU$) and cross-correlations ($TQ,\, QU,\, TU, QT,\, UQ,\, UT$). As the \lb\ scanning strategy is optimized to have redundancy in the pixel cross-linking, each $3\times3$ block of the $\mathcal{N}_{pp}^{-1}$ matrix is expected to be invertible and  almost diagonal. Any departures from symmetry are mainly  attributed to numerical round-off errors. The elements of each block are displayed as six separate maps in Figure~\ref{fig:covs}.  
 
In all the matrices, the ecliptic poles and equator are discernible due to lower correlations.  This effect is particularly noticeable in the auto-correlation fields, but it can also be distinguished in the cross-correlation fields. This pattern arises from the scanning strategy collecting more samples  at the poles and along the equator. Additionally, the $QQ$ and $UU$ matrices exhibit twice the variance of the $TT$ matrix, and the $QU$  covariances are higher than those in the $TQ$   and $TU$ cases.
Specifically, the $TQ$ and $TU$ correlations are expected to be exactly zero by construction since detector pairs are always orthogonal, causing terms like $\operatorname{Cos}(2\alpha)+\operatorname{Cos}(2\alpha+\pi)$ and $\operatorname{Sin}(2\alpha)+\operatorname{Sin}(2\alpha+\pi)$ to cancel. The small observed values ($\sim10^{-8}$) stem from numerical noise.
In contrast, QU does not vanish at the measurement level, as terms like $\operatorname{Sin}(2\alpha) \operatorname{Cos}(2\alpha) + \operatorname{Sin}(2\alpha+\pi) \operatorname{Cos}(2\alpha+\pi)$ are not vanishing by construction. While their mean goes to zero with sufficient observations, finite sampling over one year duration and at a certain resolution, leaves a small residual.

\begin{figure}[!ht]
    \centering
    \includegraphics[width=0.85\textwidth]{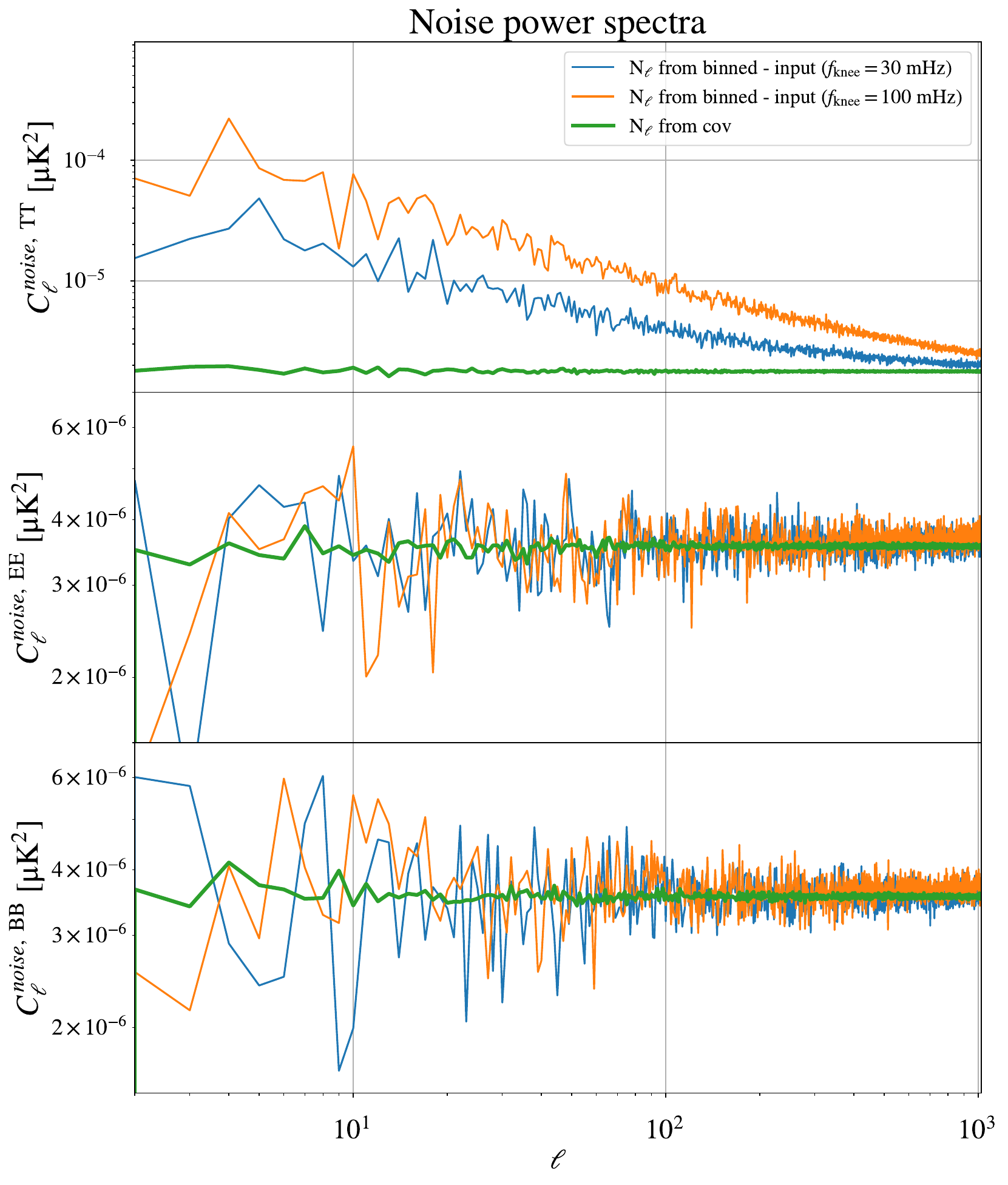}
    \caption{Noise-only power spectra from the difference between binned output and input maps (solid blue/orange) for a single simulation, alongside the average power spectra of white noise from 50 realizations (solid green) based on the noise covariance matrices in Figure~\ref{fig:covs}. Top panel: $TT$. Central panel: $EE$. Bottom panel: $BB$. The shown quantities refer to the M1-140 channel of MFT and to both the $1/f$ noise levels ($f_{\rm knee}= 30$ mHz, $f_{\rm knee}= 100$ mHz).}
    \label{fig:covs_validation_test}
\end{figure}

We present a validation test for the covariance matrix, focusing on the impact of the mapmaker on temperature and polarization, demonstrating the HWP's effectiveness in reducing the $1/f$ noise in polarization maps.
We compare the binned output maps with the input coadded maps (which include CMB, foregrounds, and $1/f$ and white noise) by taking their difference, as shown in the bottom rows of Figure~\ref{fig:maps_input_vs_output}. This subtraction isolates the noise contribution. Next, we compute the full sky power spectrum of the map differences using \healpy's \texttt{anafast} algorithm. In parallel, we generate 50 temperature and polarization maps as Gaussian realizations with a null mean and variances corresponding to the $TT,\, QQ,$ and $UU$ diagonal terms of the covariance matrices displayed in Figure~\ref{fig:covs}. Then we calculate the spectrum for each realization and compute their average. Figure~\ref{fig:covs_validation_test} compares the spectra derived from the map differences with the average spectra from the noise realizations based on the covariance matrix. Specifically, it shows the results for the M1-140 channel of MFT, comparing one simulation for both $f_{\rm knee}=30$ mHz and $f_{\rm knee}=100$ mHz to the average of 50 noise realizations.

As expected, the noise power spectra obtained from the covariance matrix-based maps (solid green line) are flat across all multipoles, representing uncorrelated white noise. In the temperature noise power spectrum, the difference between binned and input maps (solid blue/orange lines) reveals an increase in power towards lower multipoles, driven by the presence of unmitigated $1/f$ noise. However, despite this enhancement at large scales, the \lb\ data remain signal-dominated (see Figure~\ref{fig:litebird_SNR}) at low multipoles, making this effect negligible in terms of the overall signal-to-noise ratio. Conversely, for polarization, the difference between output and input (solid blue/orange lines) is consistent with a white noise power spectrum, confirming the  effective mitigation of the $1/f$ noise due to the use of a HWP.

\subsection{Power spectra}\label{sec:power_spectra}
\begin{figure}[!ht]
    \centering
    \includegraphics[width=0.85\textwidth]{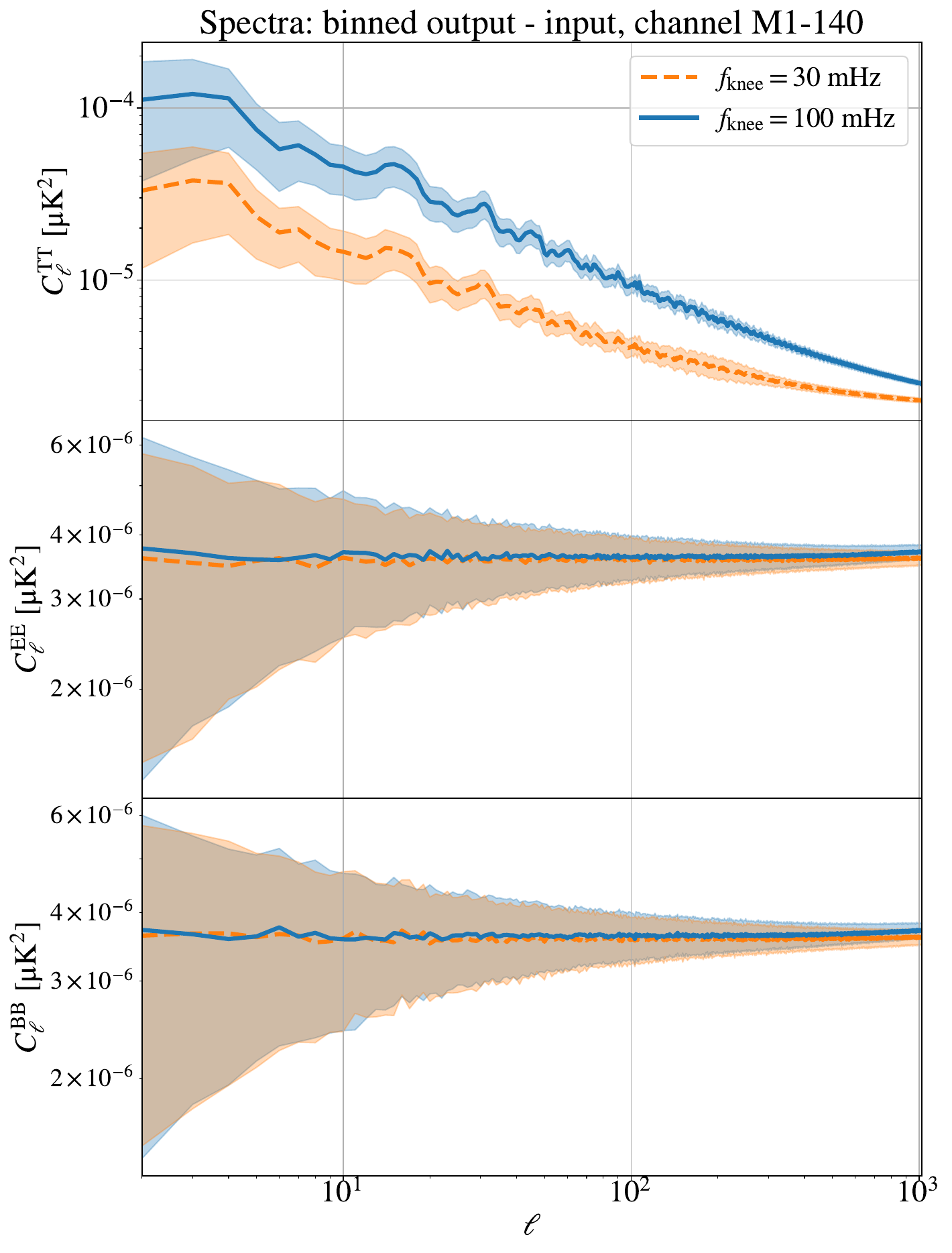}
    \caption{Full sky spectra obtained from the difference maps between the binned output maps and the input ones for the channel M1-140 of MFT. The plot shows the average of 500 realizations for both $1/f$ noise levels, with the shaded regions representing the corresponding standard deviations.}
    \label{fig:spectra_output_minus_input}
\end{figure}

% In Figure~\ref{fig:spectra_input_vs_output}, we present a comparison of the beam deconvolved power spectra derived from the input coadded maps and the output maps. The results refers to the average of 500 simulations of the M1-140 channel of MFT. To calculate the spectra, we utilized \healpy's \texttt{anafast} algorithm, considering the full sky and deconvolving for the instrumental Gaussian beam.
% \begin{figure}[!ht]
%     \centering
%     \includegraphics[width=0.85\textwidth]{fig/out_vs_in_average_of_500sims.pdf}
%     \caption{Full-sky deconvolved spectra obtained from the input coadded (black line) and output (blue/orange lines) maps for the channel M1-140 of MFT. The plot shows the average of the 500 input coadded and binned map (for both levels of the $1/f$ noise).}
%     \label{fig:spectra_input_vs_output}
% \end{figure}
% At lower multipoles, where the signal-to-noise ratio is higher, the power spectra derived from the input and output maps exhibit perfect agreement. This consistency is observed for both levels of $1/f$ noise. However, at higher multipoles, the power spectra obtained from the output maps significantly increase. This increment is attributed to the beam-deconvolved noise contribution, which is absent in the input maps. As a result, the noise component begins to dominate the power spectra at higher multipoles, leading to the observed deviations between the input and output maps.\par
In this section we present a comparison among different power spectra. To calculate the spectra, we utilized  \healpy's \texttt{anafast} algorithm, considering the full sky. Figure~\ref{fig:spectra_output_minus_input} displays the power spectra derived from the differences between the output maps and the input coadded maps. The results refer to the average of 500 simulations of the M1-140 channel of MFT. 
% The instrumental Gaussian beam has been deconvolved.
As anticipated, the temperature noise spectrum with $f_{\rm knee}=$ 100 mHz is larger than with $f_{\rm knee}=$ 30 mHz. This outcome aligns with expectations, as higher $f_{\rm knee}$ values correspond to increased noise levels in the simulations. Regarding the polarization case, the HWP demonstrates its effectiveness in reducing $1/f$ noise for both knee frequencies as the noise curves virtually overlap.

\subsection{Computational cost}\label{sec:computational_cost}
The simulations have been produced at the computing facilities at the MARCONI-CINECA in Italy. The production involved 370 computing nodes,  accounting for 48 Intel Xeon 8160 (SkyLake) cores per node. The total computational cost for the simulation production and validation is estimated to be  600 thousand CPU-h. The outputs, described in Appendix~\ref{sec:products}, encode 500 realizations of maps with one single realization of TODs occupying a total of about  35 TB of disk space. TODs have been currently stored in two different facilities to enable data retrieval and reproducibility  for future studies. 

We predict the production of full focal plane simulations within the next few years, with an estimated computational cost of 1 million CPU-h. This estimate is derived by scaling the cost of current simulations, assuming a consistent number of Monte Carlo realizations.
%To produce the first simulation, i.e. simulation 0, together with the tests of the pipeline and the validation of the first results, we employed 290~kCPUh. To produce the simulations from 1 to 49, in addition to simulation 0 which provides both binned and destriped maps, a total of 280~kCPUh were utilized. Finally, for simulations from 50 to 199, for which we run only the binner mapmaker, we exploited \Marco{Luca could you write the value?}~kCPUh. This number is considerably smaller than that for the preceding 49 simulations due to the significantly higher computational demands of the destriper compared to the binner mapmaker.\par

%\peppe{i would cross-out the paragraph above and just quote an overall cumulative number of CPUh and also refer in which system they have been performed. }
%The 200 input CMB, 200 input foreground, 200 output binned and 50 output destriped maps for the 22 channels totally occupy 900 GB of disk space. The 22 noise covariance matrices take 5 GB of disk space. The TOD for the first simulation, for the 22 channels and for the different components (CMB, foreground, dipole, white noise, $1/f$ + white noise with $f_{\rm knee}=30$ mHz, 1/f + white noise with $f_{\rm knee}=100$ mHz) together with the pointing information (colatitude, longitude, polarization angle) occupy a total of 34 TB.

\section{Conclusions}\label{sec:conclusions}
This paper presents the first release of the official \lb\ simulations. Through the scanning of input convolved CMB and foreground maps, along with the incorporation of white noise, two types of $1/f$ noise (with $f_{\rm knee}=30$ and 100 mHz), and the dipole signal, we simulated one year of Time Ordered Data (TOD) for approximately one-third of \lb's total detectors. To ensure fidelity, the white noise level was carefully rescaled to match the baseline three-year mission and the entire focal plane. The output comprises 500 binned maps and a noise covariance matrix for each of \lb's 22 channels. Additionally, we saved TOD and pointing information for the first simulation, for each of the 22 channels and for each component separately. Moreover, the TODs presented in this work have been used within the Cosmoglobe framework and results will be presented in \cite{Aurlien_2025}.\par
Notably, our findings demonstrate that the (ideal) HWP effectively mitigates $1/f$ noise (see Figs.~\ref{fig:covs_validation_test},~\ref{fig:spectra_output_minus_input}), as evidenced by the virtually flat spectra for polarization. \par
For the next rounds of simulations, we plan to increase the number of detectors considered, thus producing full focal plane simulations and extend the simulation time to the baseline three-year mission. Moreover, our focus will shift towards incorporating several systematic effects, beyond the existing $1/f$ noise currently employed in this simulation effort. These additional systematics encompass a wide range of considerations, such as gain drifts, downtime occurrences, impacts of cosmic rays, as well as beam and HWP systematics. We aim at  describing in  a publication in the coming year, the simulations encoding these systematics effects that are being produced at the time of writing this work.
% To effectively disentangle and analyze diverse effects, we plan to implement various combinations of data splits, i.e., including both temporal and detector splits.

\acknowledgments
\textit{LiteBIRD} (phase A) activities are supported by the following funding sources: ISAS/JAXA, MEXT, JSPS, KEK (Japan); CSA (Canada); CNES, CNRS, CEA (France);
DFG (Germany); ASI, INFN, INAF (Italy); RCN (Norway); MCIN/AEI, CDTI (Spain); SNSA, SRC (Sweden); UKSA (UK); and NASA, DOE (USA).

This work has received funding by the European Union’s Horizon 2020 research and innovation program under grant agreement no. 101007633 CMB-Inflate. We further acknowledge financial support under the National Recovery and Resilience Plan (NRRP), Mission 4, Component 2, Investment 1.1, Call for tender No. 104 published on 2.2.2022 by the Italian Ministry of University and Research (MUR), funded by the European Union – NextGenerationEU– Project Title ”SHIFT” – CUP 55723062008 - Grant Assignment Decree No. 962 adopted on June 30th 2023 by the Italian Ministry of Ministry of University and Research (MUR).

We acknowledge the use of \texttt{numpy}~\cite{Harris:2020xlr} and \texttt{matplotlib}~\cite{Hunter:2007ouj} software packages, and the use of computing facilities at CINECA. Some of the results in this paper have been derived using the \texttt{healpy}~\cite{Zonca2019} and \healpix~\cite{Gorski:2004by} packages.

\bibliographystyle{aa}

\bibliography{bibliography,Planck_bib}

\appendix
\section{Products of the pipeline}\label{sec:products}
We summarize all the pipeline products in Table~\ref{tab:all_products}. As already stated in Section~\ref{sec:computational_cost}, for the first realization of simulations, we generated six separate TODs for each channel: \texttt{tod\_cmb}, \texttt{tod\_fg}, \texttt{tod\_wn}, \texttt{tod\_wn\_1f\_100mHz}, \texttt{tod\_wn\_1f\_30mHz}, and \texttt{tod\_dip}. Each TOD contains only the specific component indicated by its name. Additionally, a noise covariance matrix in pixel space was produced for each channel, computed only once, as it is independent of the specifics of each simulation (see Eq.~\ref{eq: mapmaking}). Finally, we generated a total of 1000 binned maps per channel: 500 coadded maps at the \texttt{nside=512 } pixel resolution, with $f_{\rm knee} = 30$ mHz and 500 coadded maps with $f_{\rm knee} = 100$ mHz. Each coadded map includes contributions from the CMB, foregrounds, $1/f$ noise, and white noise.
\begin{table}[ht]
    \centering
    \begin{tabular}{c}
    \hline
    Products per channel\\
    \hline\hline
    \texttt{tod\_cmb} \\ \texttt{tod\_fg} \\ \texttt{tod\_wn} \\ \texttt{tod\_wn\_1f\_100mHz} \\ \texttt{tod\_wn\_1f\_30mHz} \\ \texttt{tod\_dip} \\
    \hline
    Covariance \\
    \hline
    500 coadded maps ($f_{\rm knee} = 30$ mHz) \\
    500 coadded maps ($f_{\rm knee} = 100$ mHz) \\
    \hline
    \end{tabular}
    \caption{For the 500 simulations, we generated 500 coadded maps, including the CMB, foregrounds (dust, synchrotron, free-free and Galactic CO emission), $1/f$ noise and white noise, by binning the observations for both $1/f$ noise knee frequencies. For the first simulation only, we also saved the TOD components and the noise covariances in pixel space to disk.}
    \label{tab:all_products}
\end{table}
\section{Comparison between input and output maps}\label{sec:maps}
\begin{figure}[!ht]
    \centering
    \includegraphics[width=.85\textwidth]{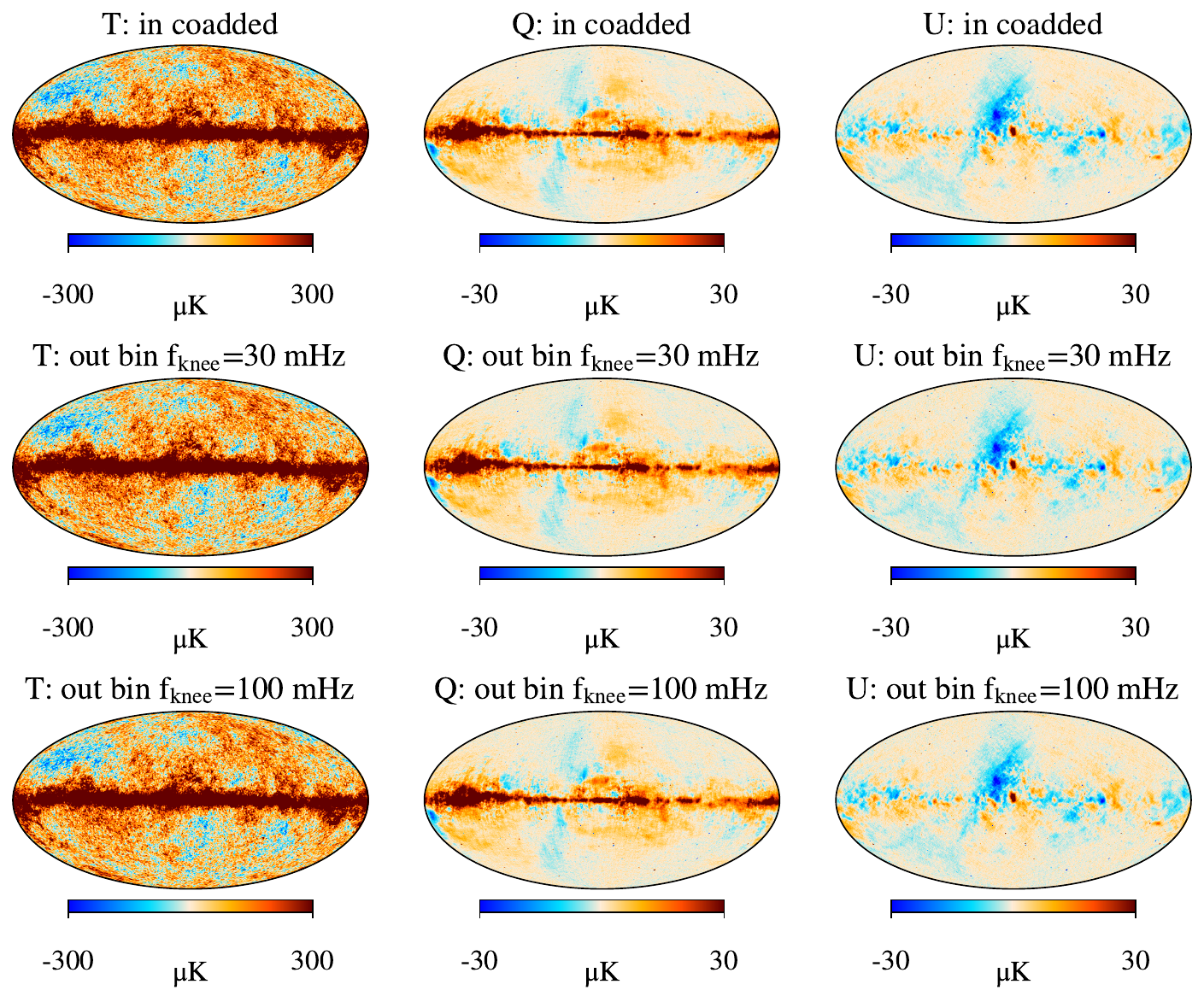}
    \includegraphics[width=.85\textwidth]{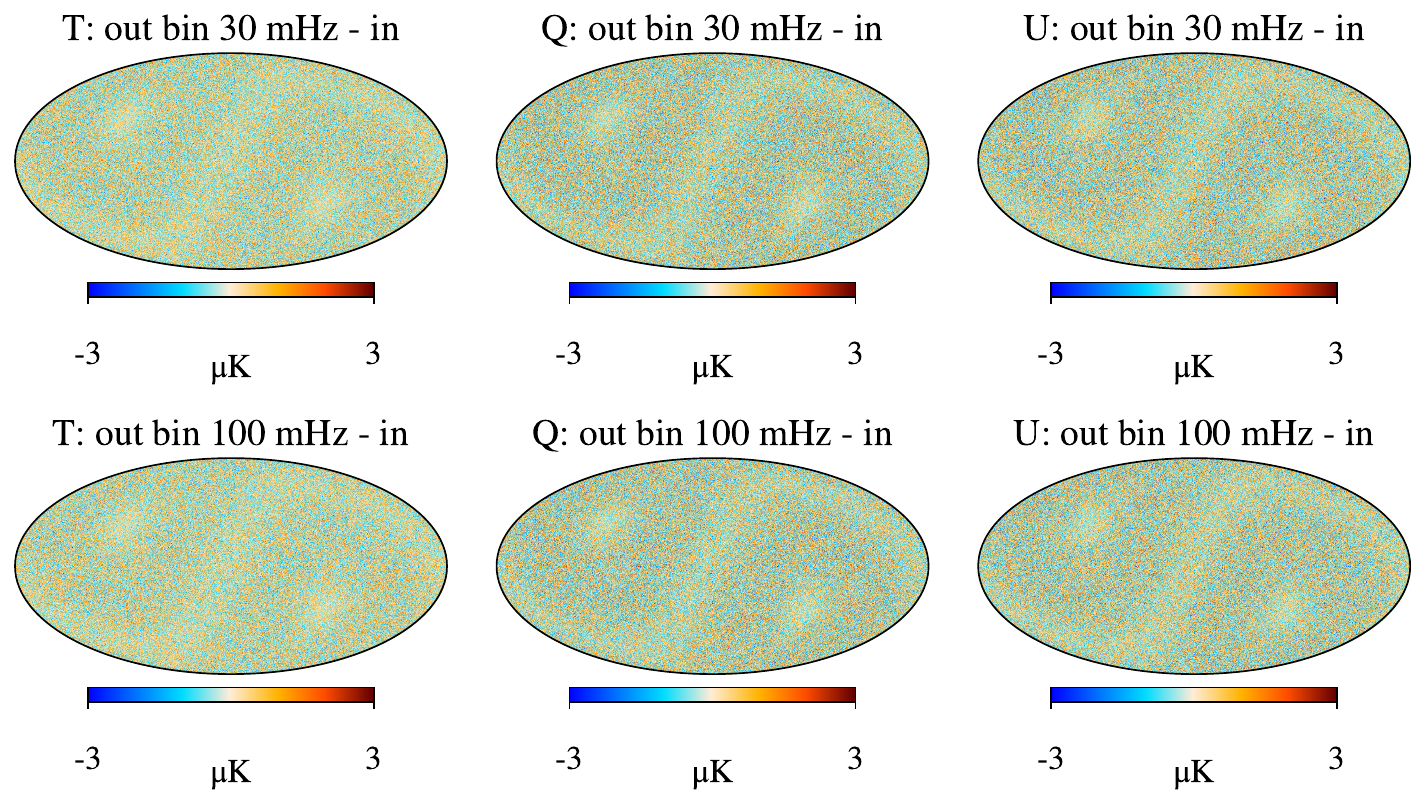}
    \caption{Comparison between input and output maps for the M1-140 channel of MFT. Left column: $T$. Central column: $Q$. Right column: $U$. From top to bottom rows: 
    input coadded (CMB + foregrounds); 
    binned output coadded + white noise + $1/f$ noise with $f_{\rm knee}=100\ \mathrm{mHz}$; 
    binned output coadded + white noise + $1/f$ noise with $f_{\rm knee}=30\ \mathrm{mHz}$;
    binned output (CMB + fg + white noise + $1/f$ noise with $f_{\rm knee}=30\ \mathrm{mHz}$) minus coadded input; binned output (CMB + fg + white noise + $1/f$ noise with $f_{\rm knee}=100\ \mathrm{mHz}$) minus coadded input.}
    \label{fig:maps_input_vs_output}
\end{figure}

Here, we present a comparison between the output and input maps. Figure~\ref{fig:maps_input_vs_output} shows the maps produced by the simulation pipeline alongside the input maps, for the first simulation and the M1-140 channel of MFT. Additionally, the fourth and fifth rows display the differences between the output and input maps.
The output maps, when compared to their input counterparts (Figure~\ref{fig:maps_input_vs_output}) and their differences, do not reveal any notable structures, confirming an accurate scanning of the input maps. The anisotropies observed in the two bottom rows primarily stem from the presence of noise.

\end{document}